\documentclass[aps,prl,twocolumn,superscriptaddress, showpacs]{revtex4-1}
\usepackage{amssymb}
\usepackage{graphicx}
\usepackage{amsmath}
\usepackage{amsfonts}
\usepackage{amssymb}

\begin{document}
\title{How to make graphene superconducting}

\author{Gianni Profeta}
\affiliation{SPIN-CNR - Dipartimento di Fisica Universit\'a degli Studi di L'Aquila, I-67100 L'Aquila, Italy}
\author{Matteo Calandra}
\affiliation{IMPMC, Universit Paris 6, CNRS, 4 Pl. Jussieu, 75015 Paris, France}
\author{Francesco Mauri}
\affiliation{IMPMC, Universit Paris 6, CNRS, 4 Pl. Jussieu, 75015 Paris, France}


\begin{abstract}

\end{abstract}

\maketitle

Graphene \cite{novoselov} is the physical realization
of many fundamental concepts and phenomena 
in solid state-physics\cite{GeimReview}, but
in the long list of graphene remarkable properties
\cite{Katsnelson,novoselovQHE,Zhang,Nair},
a fundamental block is missing: superconductivity.
Making graphene superconducting is relevant as the easy manipulation
of this material by nanolytographic techniques paves the way to nanosquids,
one-electron superconductor-quantum dot devices\cite{DeFranceschi,Huefner}, 
superconducting transistors at the nano-scale\cite{Delahaye} and cryogenic solid-state coolers\cite{sara}.

Here we explore the doping of graphene by adatoms coverage.
We show that the occurrence of superconductivity depends on the
adatom in analogy with graphite intercalated compounds (GICs).
However, most surprisingly, and contrary to the GIC case\cite{Genevieve,weller}, Li covered graphene is 
superconducting at much higher temperature with respect to  Ca covered graphene. 

As graphene itself is not superconducting, phonon-mediated superconductivity 
must be induced by an enhancement of the electron-phonon coupling
$(\lambda)$,
\begin{equation}
\lambda=\frac{N(0) D^2}{M \omega_{ph}^2}
\label{eq:Hopfield}
\end{equation}
In Eq. \ref{eq:Hopfield} $N(0)$ is the electronic density of states (DOS) at the 
Fermi level, $D$ is the deformation potential, while $M$ and $\omega_{ph}$ 
are effective atom mass and phonon frequency 
that in metallic alloys reflect the role of the different atomic species and
phonon vibrations involved in superconductivity. 
In undoped graphene $\lambda$ is small and phonon-mediated superconductivity
does not occur as the small number of carriers, intrinsic in a semimetal,
leads to a vanishingly small $N(0)$.
In this respect the situation is similar to the bulk graphite case, where,
without intercalation of foreign atoms superconductivity is not stabilized.

A first guess to induce superconductivity could then be to fill by
rigid-band doping the carbon $\pi-$states in order to have enough carriers.
However, beside the fact that the $\pi-$ DOS grows very slowly with doping,
this hurts against two major difficulties.
First, even if the deformation
potential related to coupling between $\pi-$bands and in-plane phonon vibrations
is very large and leads to Kohn anomalies \cite{Piscanec},  
these vibrations are highly 
energetic ($\omega_{ph}\approx 0.16 $ eV) and
$\lambda$ is suppressed by the $\omega_{ph}^2$ factor in the denominator.
Second, symmetry forbids the coupling between $\pi-$states 
and the softer out-of-planes vibrations.

\begin{figure}
\includegraphics[scale=0.3,angle=0]{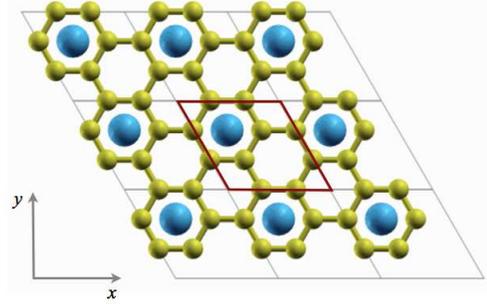}
\caption{ Crystal Structure of metal adatom covered graphene: yellow (blue) spheres represent carbon (metal) atoms. 
Adatoms sit on the hollow site of graphene layer.}
\label{crystal}
\end{figure}

In order to promote  coupling to carbon out-of-plane vibrations
it is then necessary to promote new electronic states at the Fermi level
as it happens in GICs.
Indeed in superconducting GICs an intercalant bands
(interlayer state) occurs at the Fermi level \cite{Csanyi} having multiple beneficial
effects on $\lambda$ as (i) the number of carriers is enhanced,
(ii) coupling to C out-of-plane vibrations is promoted, (iii) coupling
to intercalant vibrations occurs with a corresponding enhancement of the deformation
potential and a reduction of the effective $M\omega_{ph}^2$ term in the
denominator of Eq. \ref{eq:Hopfield}. 

In GICs not all kind of intercalant atoms leading to an interlayer
state are equally effective in increasing $T_c$ \cite{calandra3}. 
The larger T$_c$ is indeed obtained when the distance ($h$) between the intercalant
atom and the graphite plane is smaller. 
The reason is that the closer the intercalant electrons are to the 
planes, the larger are the carbon out-of-planes and the intercalant-modes deformation potentials $D$.
This is demonstrated by the increase of 
T$_c$ from BaC$_6$ (not superconducting) to SrC$_6$ 
(T$_c=1.65$ K)\cite{Boeri} and finally to CaC$_6$ (T$_c=11.5$ K)\cite{weller,Genevieve,footnote} 
and by the increase of CaC$_6$ critical temperature under hydrostatic pressure\cite{Gauzzi}.

However a too small intercalant-graphite layer distance could also be detrimental for superconductivity.
Indeed the quantum-confinement of the interlayer state in a too narrow region between the planes could result
in a upshift of the intercalant band well above the Fermi energy. In this case the ionization of the
intercalant atom is complete and
superconductivity is totally suppressed,  as in bulk LiC$_6$\cite{Csanyi}.

\begin{table}
\caption{ Calculated physical parameters: Optimized structural parameters (in \AA), electron-phonon coupling ($\lambda$), logarithmic frequency average ($\omega_{log}$ in cm$^{-1}$) and superconducting critical temperature (T$_c$ in Kelvin) for CaC$_6$ and LiC$_6$ systems. }
\centering 
\begin{tabular}{c c c c c c }
\hline
                                &  $a$      &      $ h$      &   $\lambda$   &    $\omega_{log}$   &   T$_c$  \\
\hline
CaC$_6$ bulk         &  4.30     &      2.19      &    0.68            &     284.3                  &   11.5     \\
CaC$_6$ mono      &   4.26     &      2.24      &    0.40            &    309.9                   &  1.4       \\
\hline\hline
LiC$_6$ bulk           &  4.29      &     1.83       &  0.33              &     715.7                  &  0.9       \\
LiC$_6$ mono         & 4.26      &      1.83      &   0.61             &      277.8                 &   8.1       \\
\hline

\end{tabular} 
\label{table} 
\end{table}

Following these accepted guidelines for graphite intercalated compounds,
we explore the doping of graphene by foreign-atoms coverage (see Fig. \ref{crystal}) and the possibility in  in generating superconductivity in graphene  by first-principles density functional theory calculations\cite{qe, pz}.

The parallel with GICs  is tempting, and in particular, considering that Calcium intercalated graphite shows the highest superconducting critical temperature among GICs, the first example we consider is Calcium doped graphene 
and for comparison
we simulated Ca  
intercalated graphite.
Calcium, as other alkaline metals, adsorbs in the hollow site of graphene (see Fig.\ref{crystal}, Table\ref{table})

\begin{figure}
\begin{tabular}{c}
\includegraphics[scale=0.4,angle=0]{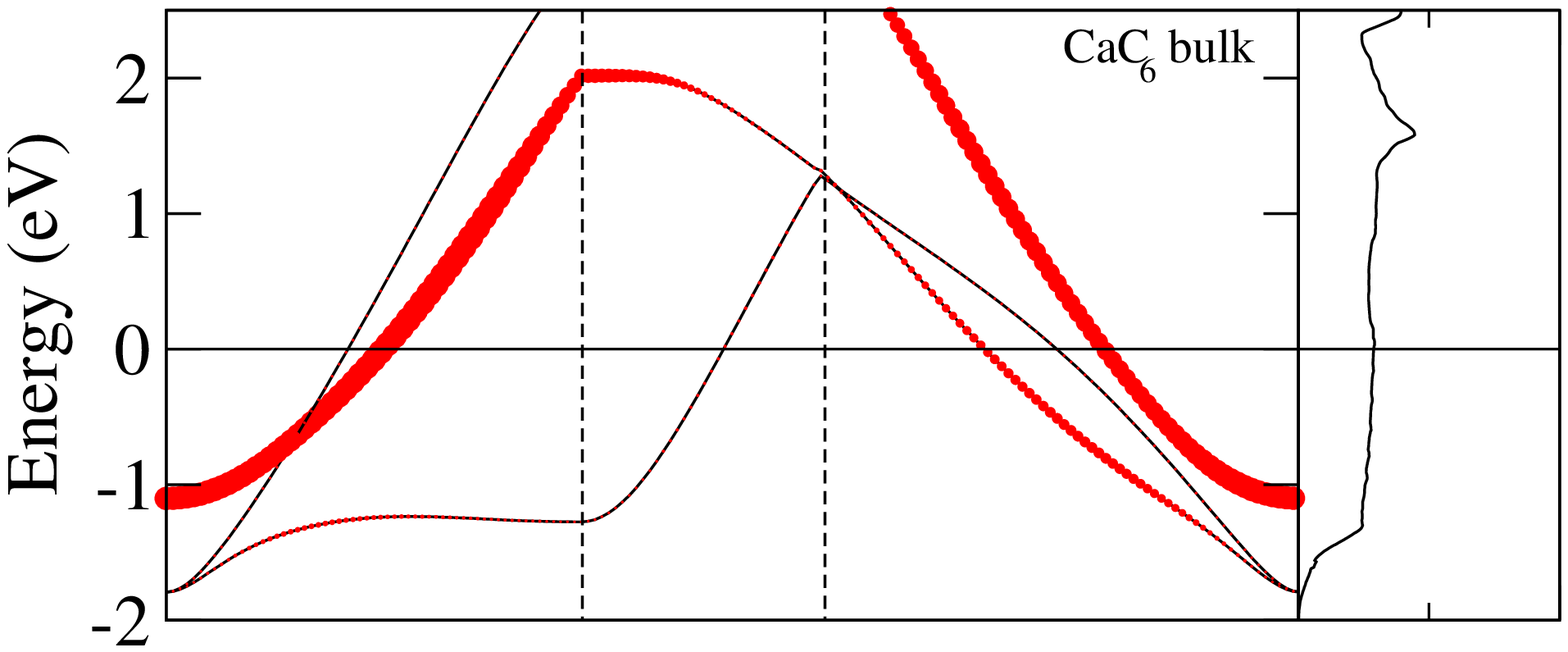}\\
\includegraphics[scale=0.4,angle=0]{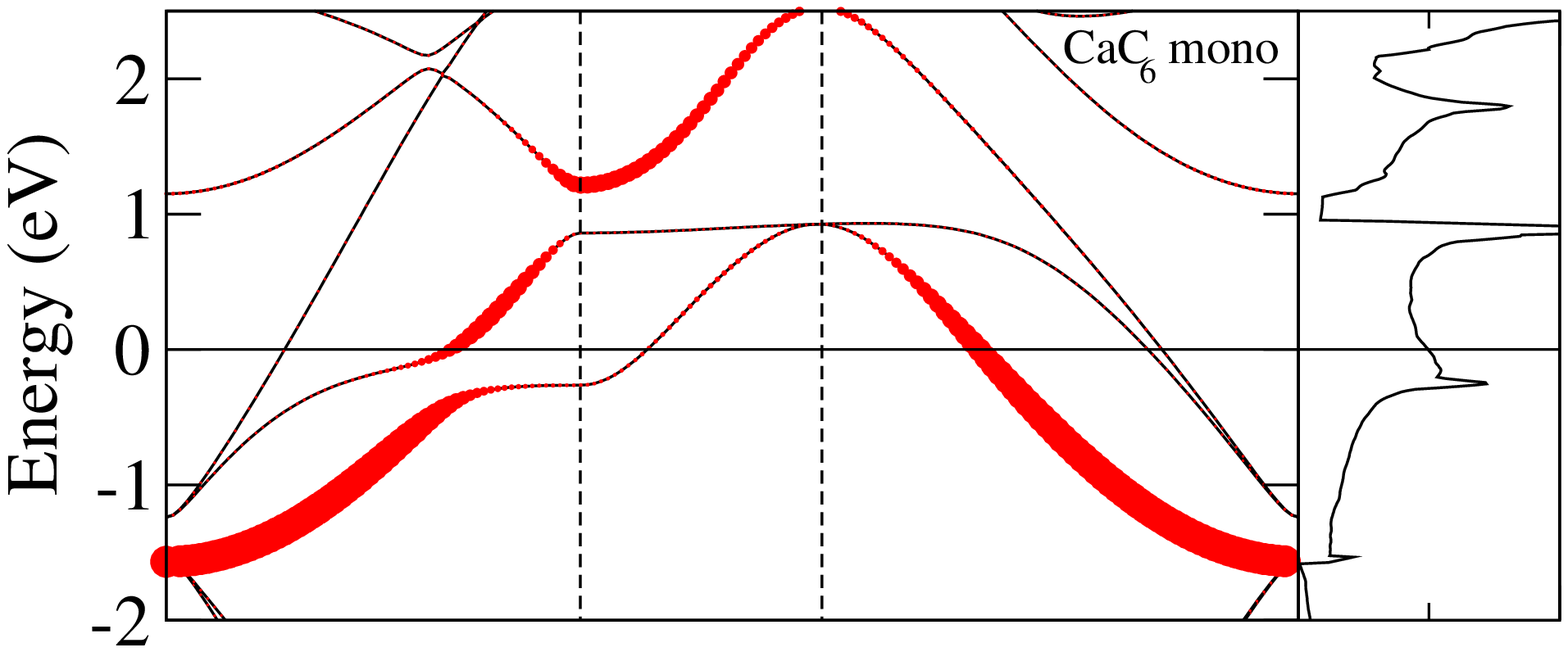}\\
\includegraphics[scale=0.4,angle=0]{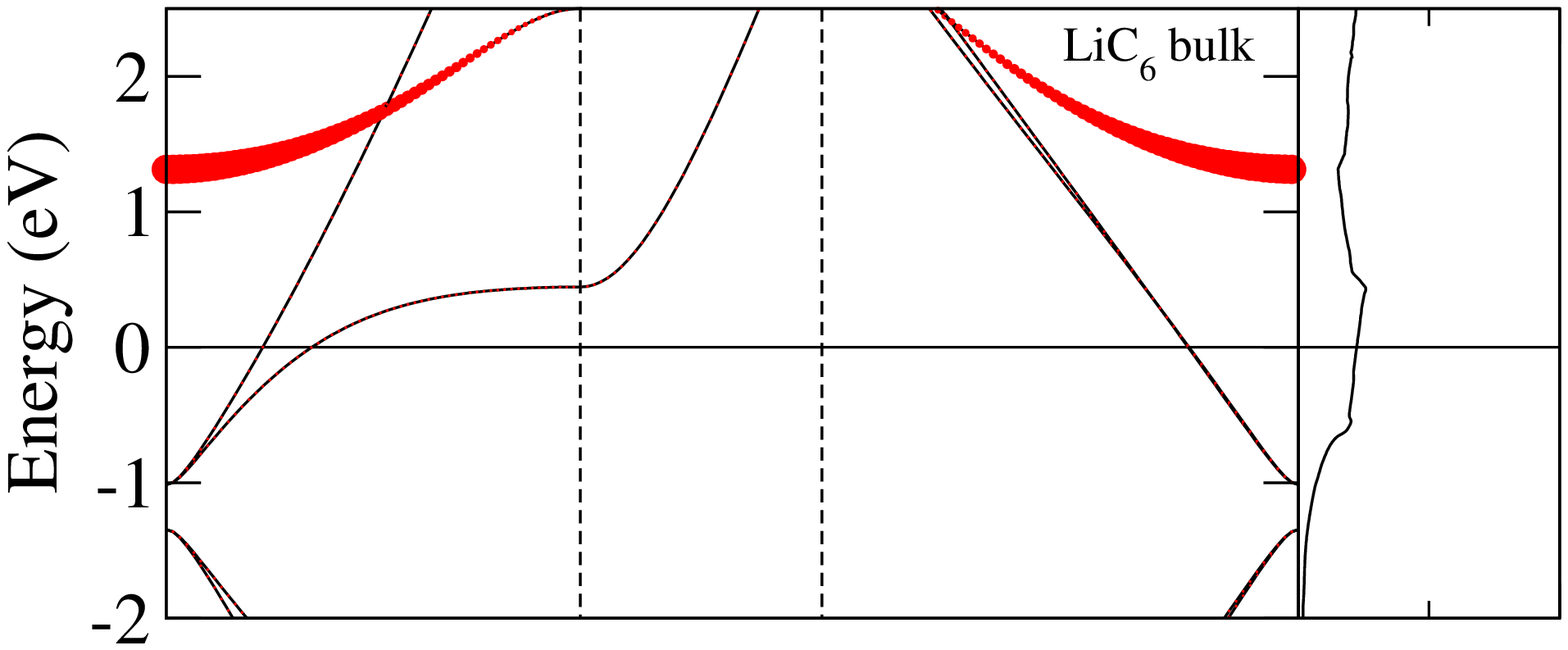}\\
\includegraphics[scale=0.4,angle=0]{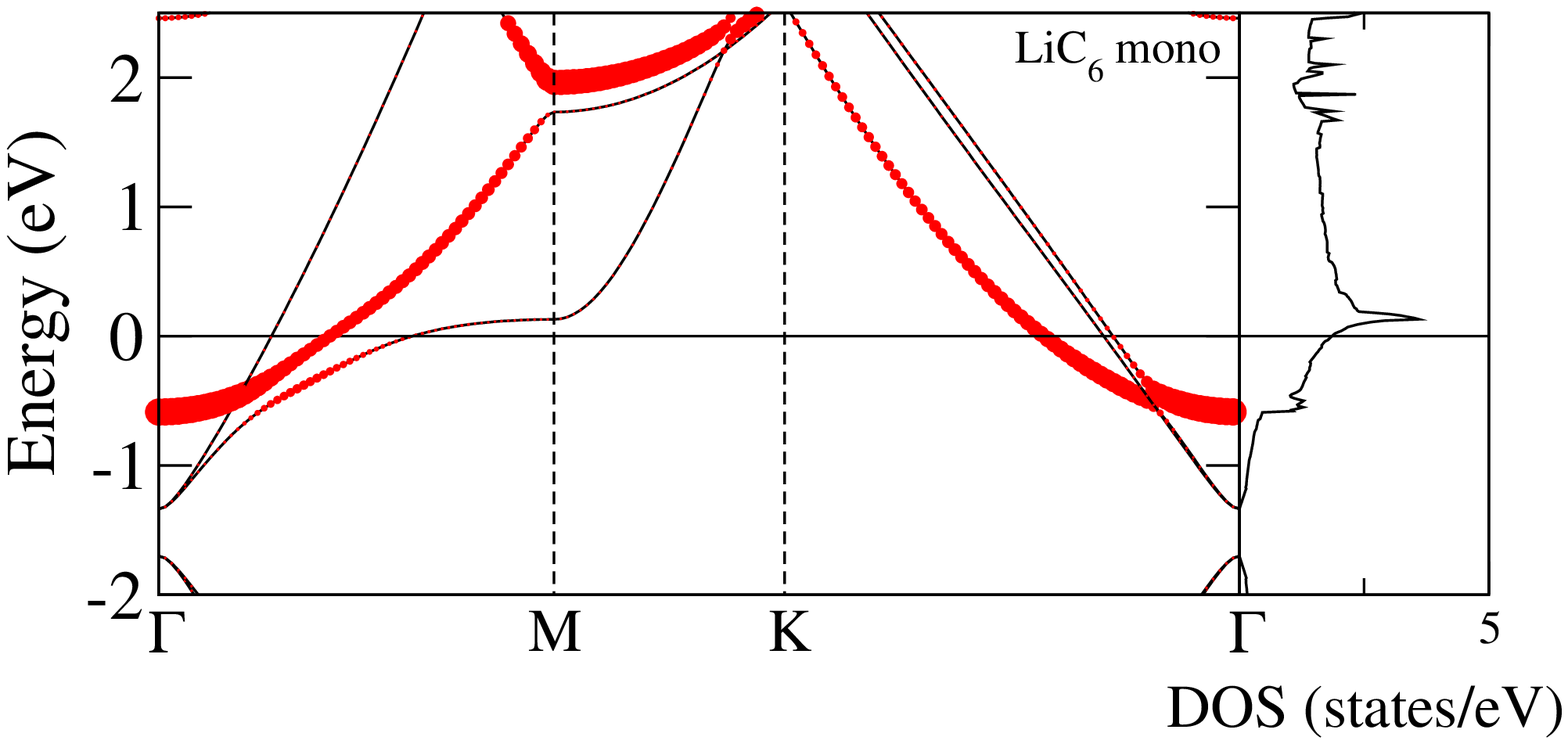}
\end{tabular}
\caption{{ Electronic band structure}: From top to bottom: band structure of CaC$_6$ bulk compound (plotted along the high symmetry lines of the monolayer lattice), 
CaC$_6$ monolayer and LiC$_6$ bulk and  monolayer.
Thickness of the bands is proportional to the intercalate/adatom $s$ character. The Fermi energy is set at zero.}
\label{fig1}
\end{figure}

The electronic band structure of both compounds is reported in Fig. \ref{fig1} and shows that the deposition of Ca on-top of graphene still preserve the interlayer band at the Fermi level present in the bulk, fundamental for the appearance of superconducting phase in GICs.
The removal of quantum confinement along the $c$-direction, due to the periodically stacked graphene layers, in passing from the bulk to the monolayer system, lowers the energy of the IL state, that will result more occupied and  carbon $\pi$-bands  less doped. 

\begin{figure}
\includegraphics[scale=0.4,angle=0]{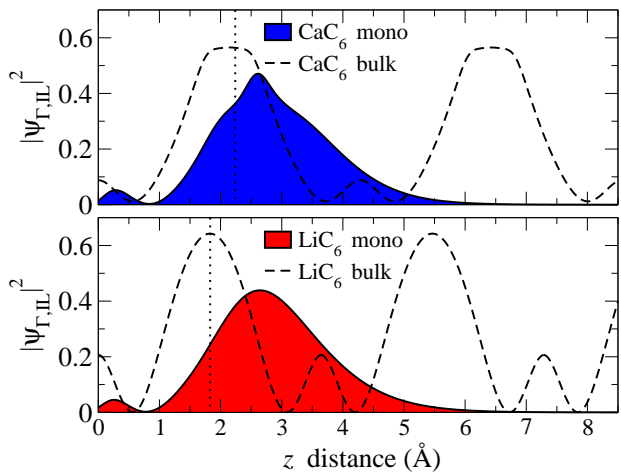}
\caption{{ Interlayer state wavefunction}: Planar (in the $xy$ direction ) average of $|\phi_{\Gamma,IL}|^2$ along the perpendicular (with respect to the graphene layer) direction ($z$). Vertical dashed lines represent $z$-position of Ca and Li adatoms. }
\label{fig2}
\end{figure}

However, the sole presence of the IL state at the Fermi level being a necessary condition, is not sufficient to  guarantee large electron-phonon coupling, unless this state is not coupled with out-of-plane vibrations of  graphene lattice. 
As the localization of the IL state as close as possible to the graphene layer\cite{Csanyi, calandra, bachelet} enhances the coupling,
 in Fig.\ref{fig2} we compare the planar-average of the IL charge density 
(calculated at the $\Gamma$-point of the Brillouin Zone) in the case of bulk and monolayer CaC$_6$. 

In the monolayer, the  IL charge density  spills-out in the vacuum region while  in the bulk case, it is much more confined in between the 
graphene and adatom layers (roughly the region from $z$=0 \AA\  to $z$=2 \AA\ ).

\begin{figure}
\begin{tabular}{c}
\includegraphics[scale=0.4,angle=0]{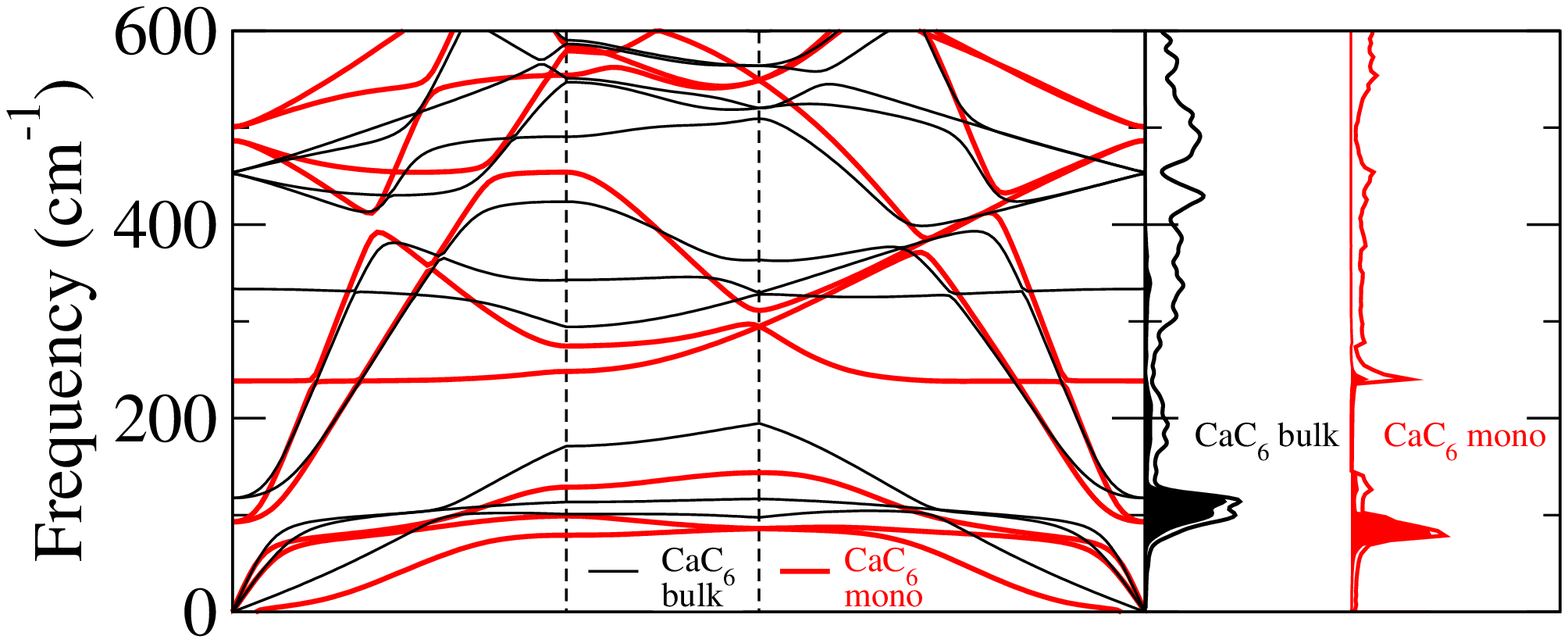}\\
\includegraphics[scale=0.4,angle=0]{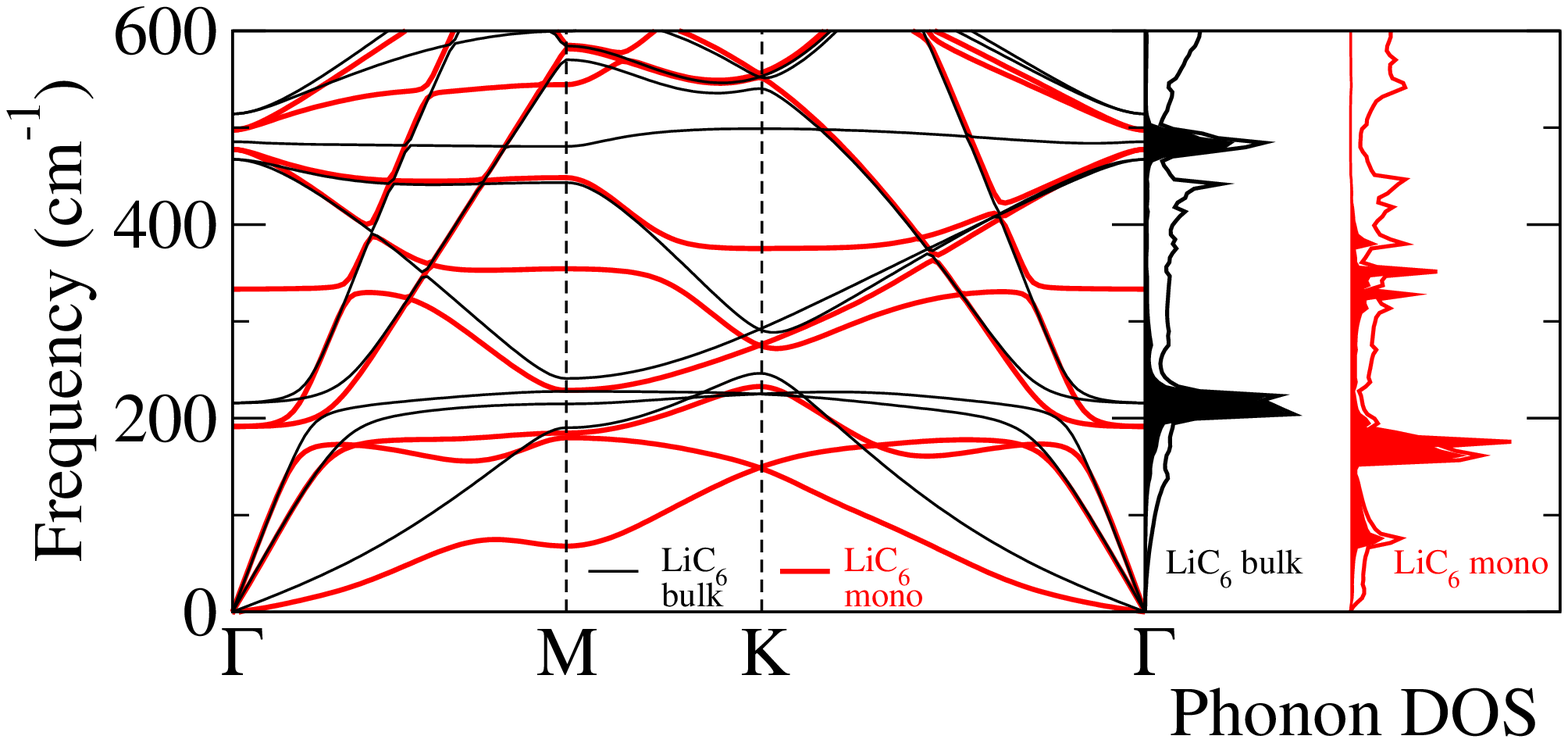}
\end{tabular}
\caption{{ Dynamical properties}: Phonon frequencies dispersion and phonon density of states (PDOS) of  CaC$_6$ (upper panel) and LiC$_6$ (lower panel). 
Bulk (black) and  monolayer (red) cases are considered resolving the  adatom (shaded)  contribution to the total (solid line) PDOS.} 
\label{fig3}
\end{figure}

\begin{figure}
\includegraphics[scale=0.53,angle=0]{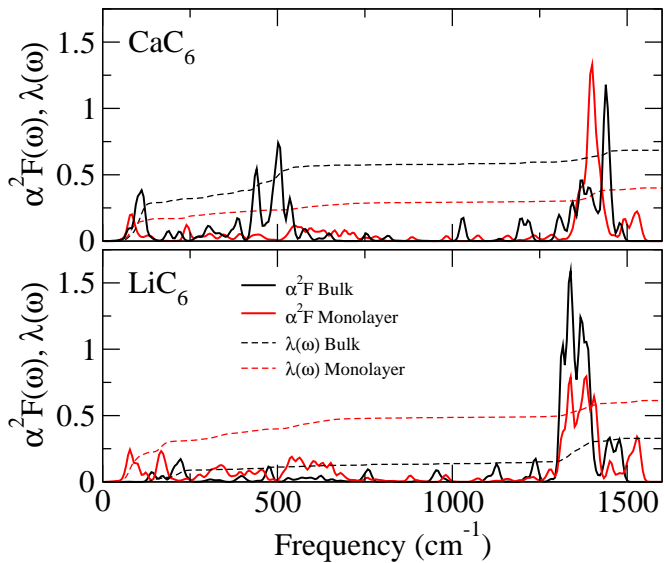}
\caption{{ Eliashberg function}: CaC$_6$ and LiC$_6$ (bulk and monolayer) $\alpha^2F(\omega)$. In the same graph is reported  $\lambda(\omega)$.}
\label{fig4}
\end{figure}

Based on this last consideration, we can infer that the superconducting critical temperature of CaC$_6$ monolayer should be  lower than its bulk counterpart.
In order to verify this hypothesis, we calculated the vibrational spectrum and the electron-phonon coupling $\lambda=2\int d\omega \alpha^2F(\omega)/\omega$  of both compounds.
The results are summarized in Table \ref{table}, Fig.\ref{fig3} and Fig.\ref{fig4}. 

First of all we notice that the monolayer system is dynamically stable, not showing the tendency to displacive instabilities, with a 
phonon dispersion characterized by three regions: 
the low energy region of adatom related modes (up to 300 cm$^{-1}$) extending up to 400  cm$^{-1}$ when mixed with 
carbon out-of-plane modes (C$_z$), 
an intermediate region of C$_z$ modes  (400-900 cm$^{-1}$)   and the high energy region characterized by  C-C stretching modes.

The major difference between  bulk and monolayer cases is the softening of Ca vibrations. 
From the inspection of the $\alpha^2F(\omega)$ (which gives the contribution of each frequency to the total electron-phonon coupling), we note 
the low contribution of the $C_z$ modes (around 500 cm$^{-1}$  to the total electron-phonon coupling in the monolayer case with respect to the bulk CaC$_6$, as expected from the above considerations.
While the critical temperature of the bulk compound is 11.5 K (see Methods) in the monolayer case it lowers down to 1.4 K.

This scenario suggests that, as a general rule, the removal of quantum confinement should be detrimental for the electron-phonon coupling when compared with respect to same bulk (three dimensional periodic) system, due to the shift of the IL wavefunction away from the graphene layer. 
Every metal covered graphene should have (at least in the same stoichiometry) a reduced superconducting critical temperature with respect to the corresponding GIC.

However, there is at least one case (to the best of our knowledge) among GICs that can be
further explored, namely stage-1 Lithium intercalated compound, LiC$_6$\cite{lic6}, 
in which the  IL state is completely empty (see Fig. \ref{fig1}), as the strong confinement along the $z$ direction (Fig.\ref{fig2})
prevents its occupation and for this reason it is not
superconducting. In this case, removal of confinement
along $c$ direction (Fig.\ref{fig2}) should bring the IL at the Fermi level,
as confirmed by the calculated band structure (Fig.\ref{fig1}).
Inspection of Fig.\ref{fig2} shows that  the spatial extension of the IL for Li  is the same as in Calcium, but being the IL strongly localized around the adatom and due its closer position to the graphene layer, we can expect an enhancing of the total electron-phonon coupling.

The phonon spectrum confirms, even in this case, the dynamical stability of the monolayer system and the comparison with the bulk counterpart reveals that  low energy adatom modes and carbon vibrations in the direction perpendicular to the plane are strongly softened in the monolayer.
This suggests that the first two phonon branches, related to the in-plane displacements of Li atoms,  and the undispersing Einstein mode  
displacing Li  and C atoms out-of-phase along the $z$-direction at around 500 cm$^{-1}$,
should undergo enhanced electron-phonon coupling with respect to the bulk. 
On the contrary, in the CaC$_6$ monolayer, adatom vibration and C vibrations along $z$ are essentially at the same energy as
in bulk CaC$_6$.

We found that bulk LiC$_6$ is  a weak electron-phonon coupling $\lambda=0.33$  superconductor with an estimated 
superconducting  critical temperature T$_c=0.9$K\cite{warningtc},  theoretically confirming the absence of superconductivity in LiC$_6$.  
In the case of monolayer LiC$_6$, the total electron-phonon coupling is $\lambda=0.61$, with a relevant enhancement of the superconducting critical temperature up to 8.1 K.
The main contributions to the electron-phonon coupling come from the low-energy Li modes and carbon vibrations along $z$, as expected from the above considerations, with a appreciable contribution  from  C-C stretching modes ($0.1$).

In summary, the ideal conditions for inducing superconductivity in graphene are
$(i)$ bring the interlayer state at the Fermi energy 
$(ii)$, localize it as close as possible to the graphene plane.
This will ensure that the presence of the interlayer state effectively switch-on the dormant electron-phonon coupling of C$_z$ modes that are inactive in the bulk. In addition, the in-plane displacement of adatoms will promote two additional scattering mechanisms, an intra-band contribution due to the 
interlayer state and an inter-band one due to the interlayer-$\pi$ scattering.
{\it Graphene can be made superconducting by the deposition
of Li atoms on the top of it.}

As Li readily intercalates into graphite even at 100 K\cite{review_ligraphite}  it is possible to incorporate Li atoms even  below the graphene sheet.  
This possibility, based on our calculations, is indeed favorable with respect to the developing of a superconducting phase.
In fact, the double adsorption should double the presence of the interlayer state at the Fermi level (one coming from each side of graphene) as we indeed verified.

Li$_2$C$_6$ shows an increased coupling in all the frequency range  due to the additional strongly coupled interlayer bands, 
with an electron-phonon coupling of 1.0 giving rise to critical temperature around 17-18 K.

We conclude noting once more that bulk LiC$_6$ undergoes a remarkable 
metal-to-superconductor transition when exfoliated to one layer. 
The reason is that in a single layer LiC$_6$ the interlayer state
cross the Fermi level, enhances the electron-phonon coupling
and induces superconductivity. Our work demonstrates
that superconducting properties of adatoms on graphene are very
different from their bulk GIC counterparts.

\section{Methods}
\footnotesize 
The results reported in the present paper were obtained from first-principles density functional theory in the local density approximation\cite{pz}.
The  QUANTUM-ESPRESSO\cite{qe} package was used with norm-conserving pseudopotentials and a plane-wave cutoff energy of 65 Ry.
All the structures considered were relaxed to their minimum energy configuration following the internal forces on atoms and the stress tensor of the unit cell.

The monolayer systems were simulated in the  $\sqrt3\times\sqrt3R30^{\circ}$ in-plane unit cell 
(with respect to standard graphene lattice,  see Fig.\ref{crystal}) with one adatom per unit cell.
Phonons frequencies were calculated in the linear-response technique on a phonon 
 wavevector  mesh of 12$\times$12 with a 14$\times$14 uniform electron momentum grid.
The electron-phonon coupling parameter was calculated with electron momentum 
$k$-mesh up to 40$\times$40

CaC$_6$ bulk compound was simulated in the experimentally found structure with 
$\alpha\beta\gamma$ stacking\cite{Genevieve} with a uniform 
electron-momentum $k$-mesh integration of $8\times 8\times 8$.
 The phonon frequencies were calculated on a $4\times 4\times 4$ 
phonon-momentum mesh and the electron-phonon coupling integrated on a  
$20\times 20\times 20$ electron-momentum mesh.
 
Bulk LiC$_6$ has an $\alpha\alpha$ stacking and was simulated with and electron-momentum 
mesh of 12$\times$12$\times$10 and $6\times 6\times 6$ phonon-momentum grid
 for the calculation of phonon frequencies.
 A 30$\times$30$\times$25 electron-momentum mesh was used for the electron-phonon 
coupling.
 
The Eliashberg function $\alpha^2F(\omega)$ is defined as:
\begin{eqnarray}
\alpha^{2}F(\omega)=\frac{1}{N(0) N_{k}} \sum_{\mathbf{kn,qm},\nu}|g_{\mathbf{kn,k+qm}}^{\nu}|^2 \times \nonumber \\
                                 \delta(\varepsilon_{\mathbf{kn}})\delta(\varepsilon_{\mathbf{k+qm}})\delta(\omega-\omega_{\mathbf{q}}^{\nu})
\end{eqnarray}
The total electron-phonon coupling $\lambda(\omega)$ plotted in Fig.\ref{fig4} is defined as: 
\begin{equation}
\lambda(\omega)=2\int_0^{\omega} d\omega' \frac{\alpha^2F(\omega')}{\omega'}
\end{equation}

The total electron-phonon coupling is $\lambda(\omega \to \infty)$.
The superconducting critical temperature was estimated using the Allen-Dynes formula with $\mu^* =0.115$, which fits the experimental critical temperature measured in CaC$_6$ GIC\cite{Genevieve}.

\acknowledgments
Work supported by: CINECA-HPC ISCRA grant,  by the EU DEISA-SUPERMAG project
and by a HPC grant at CASPUR.  Part of the calculation were performed
at the IDRIS supercomputing center (project 91202).


\begin{thebibliography} {99}
\bibitem{footnote} The distance from the graphene plane can be correlated with the ionic radii $r$. For  the alkalis and alkaline-earths 
considered thery are $r({\rm Li})=0.76$\AA\ , $r({\rm Ca})=1.0$\AA\ , $r({\rm Sr})=1.18$\AA\ , $r({\rm Ba})=1.35$\AA\ .
\bibitem{warningtc} The estimated critical temperatures for low $\lambda$ values are affected by an exponentially increasing error. Within the scope of the present paper, calculated  T$_c$ lower than 1K can be considered as 0K.
\end{thebibliography}
\end{document}